\documentclass[twocolumn,10pt]{article}

 \topmargin    - 12pt 

 \usepackage[square,numbers,sort&compress,comma]{natbib}

 \usepackage{amsmath}
 \usepackage{amssymb}
 \usepackage{caption}
 \usepackage{graphicx}
 \usepackage{latexsym}
\usepackage{psfrag}
 \usepackage{times}
 \usepackage{supertabular}

 \renewenvironment{abstract}%
              {
               \small
               {\bfseries \abstractname}
               \par
               \vspace{10pt}
              }

 \renewcommand\abstractname{Abstract}

   \newcommand{\nomenclature}
              [1]
              {
               \bgroup
               \flushleft
               \small\bf
               #1
               \par
               \egroup
              }

 \renewcommand{\section}
              [1]
              {
               \bgroup
               \flushleft
               \small\bf
               \stepcounter{section}
               \arabic{section}. #1
               \par
               \egroup
              }

 \renewcommand{\subsection}
              [1]
              {
               \bgroup
               \flushleft
               \small\em
               \stepcounter{subsection}
               \arabic{section}.
               \arabic{subsection}. #1
               \par
               \egroup
              }

 \renewcommand{\subsubsection}
              [1]
              {
               \bgroup
               \flushleft
               \small\em
               \stepcounter{subsubsection}
               \arabic{section}.
               \arabic{subsection}.
               \arabic{subsubsection}. #1
               \par
               \egroup
              }

   \newcommand{\acknowledgement}
              [1]
              {
               \bgroup
               \flushleft
               \small\bf
               #1
               \par
               \egroup
              }

   \newcommand{\sectionbib}
              [1]
              {
               \bgroup
               \flushleft
               \small\bf
               #1
               \par
               \egroup
              }

 \newcommand{\D}[2]{\frac{\partial #1}{\partial #2}}
 \newcommand{\DD}[2]{\frac{\partial^2 #1}{\partial #2^2}}

\title{\LARGE Experiments in a novel quasi-1D diffusion\\
              flame with variable bulk flow}

\author{{\large E. Robert and P.A. Monkewitz}\\[10pt]
{\footnotesize \em Laboratory of Fluid Mechanics (LMF), Swiss Federal Institute of Technology Lausanne (EPFL), CH-1015 Lausanne, Switzerland}}

\date{}

\begin{document}


\small
\baselineskip 10pt


\twocolumn[\begin{@twocolumnfalse}
\addvspace{50pt}
\maketitle
\vspace{40pt}
\rule{\textwidth}{1pt}
\begin{abstract}
%
  The novel species injector  of a recently developed research burner,
  consisting  of  an array  of  hypodermic  needles,  which allows  to
  produce quasi  one-dimensional unstrained diffusion  flames has been
  improved.  It is used in a new symmetric design with fuel \emph{and}
  oxidizer   injected   through   needle   arrays  which   allows   to
  independently choose  both the magnitude  and direction of  the bulk
  flow through the flame. A simplified theoretical model for the flame
  position with variable bulk flow is presented which accounts for the
  transport  properties  of both  reactants.   The  model results  are
  compared to experiments with  a $CO_2$-diluted $H_2$-$O_2$ flame and
  variable bulk  flow. The mixture composition  throughout the burning
  chamber   is  monitored   by  mass   spectrometry.    The  resulting
  concentration profiles  are also  compared to the  simplified theory
  and demonstrate  that the new  burner configuration produces  a good
  approximation of  the 1-D chambered diffusion flame,  which has been
  used  extensively for  the stability  analysis of  diffusion flames.
  Hence, the  new research burner  opens up new possibilities  for the
  experimental  validation  of  theoretical  models developed  in  the
  idealized   unstrained  1-D   chambered   flame  configuration,   in
  particular models  concerning the effect of bulk  flow magnitude and
  direction on flame stability. Some preliminary results are presented
  on  the  effect of  bulk  flow  direction  on the  thermal-diffusive
  cellular flame instability.
\end{abstract}
\vspace{10pt}
\parbox{1.0\textwidth}{\footnotesize {\em  Keywords:} Diffusion flame,
  unstrained, instability}
\vspace{30pt}
\rule{\textwidth}{1pt}
\vspace{40pt}
\end{@twocolumnfalse}]


 \section{Nomenclature} \addvspace{10pt}

 \begin{supertabular}{l l}
 $C_p$ & Specific heat at constant pressure\\
 $D_i$ & Diffusivity of species $i$\\
 $Le_i=\kappa/D_i$ & Lewis number of species $i$\\
 $q$ & Heat released per unit mass of fuel\\
 $Q$ & Total heat released\\
 $T_f$ & Adiabatic flame temperature\\
 $U$ & Bulk flow velocity\\
 $\bar W$ & Mean molecular weight\\
 $W_i$ & Molecular weight of specie $i$\\
 $X_f$ & Fuel mass fraction\\
 $X_o$ & Oxidizer mass fraction\\
 $x$ & Coordinate along burner length\\
 $\kappa=\lambda/(\rho C_p)$ & Thermal diffusivity\\
 $\nu_i$ & Stoichiometric coefficient of species $i$\\
 $\rho$ & Density\\
 $\phi$ & Equivalence ratio\\
 $\omega$ & Chemical reaction rate\\
 \end{supertabular}

\section{Introduction} \addvspace{10pt}

In chemically  reacting flows, instabilities arise  from the competing
mechanisms  of  thermal and  mass  diffusion.  These  thermo-diffusive
instabilities  can manifest  themselves  by cellular  patterns in  the
reaction  zone  \cite{Kim1996,  Kim1997a, Cheatham2000,  Metzener2006}
which have  been observed  in non-premixed systems.   In technological
applications,  they where  found to  play a  significant role  in soot
formation and extinction dynamics \cite{Kim1997}.

\begin{figure*}[ht!]
\centering
\includegraphics[width=144mm]{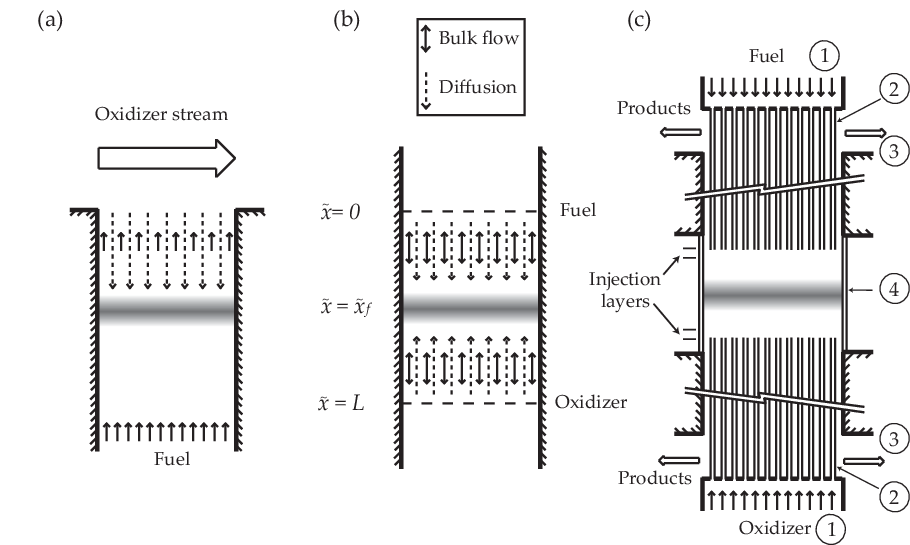}
\caption{a)  Conventional  chambered  diffusion flame.   b)  Symmetric
  chambered   diffusion   flame  with   reversible   bulk  flow.    c)
  Experimental configuration  used.  1, Injection Plenum;  2, Array of
  hypodermic injection needles; 3,  The products are allowed to escape
  between the injection tubes; 4, Quartz-walled reaction chamber.}
\label{fig:1DBurners}
\end{figure*}

Since its  introduction by  Kirkby and Schmitz  \cite{Kirkby1966}, the
idealized  chambered flame  model has  been and  continues to  be used
extensively to study the stability  of diffusion flames in the absence
of   hydrodynamic  effects   \cite{Kim1996,   Kim1997a,  Cheatham2000,
  Metzener2006,  Law2006,  Glassman1996}.    The  simplicity  of  this
configuration, illustrated in figure \ref{fig:1DBurners}a), allows for
analytical  solutions to  the  governing equations  of the  chemically
reacting  flow.   Models  developed  in this  configuration  have  the
advantage  of  uncoupling   thermo-diffusive  instabilities  from  any
hydrodynamics.   However, until  recently these  models  have remained
without experimental  validation because of the  difficulty to realize
such configurations experimentally.

The challenge  arises from the  necessity to supply the  reactants and
remove the products evenly across the burner cross section to maintain
the one-dimensional character of the  flame and avoid strain which has
a   strong  effect   on  flame   stability  and   extinction  dynamics
\cite{Chen1991}.  A  novel research burner  configuration has recently
been  introduced   that  creates  such   a  flame  \cite{LoJacono2005,
  Robert2007} by  using an array  of hypodermic needles to  supply the
reactants and allowing  the products to escape between  them.  In this
paper, a new improved symmetric version of this burner is presented in
which both reactants are supplied through hypodermic needle arrays and
the products can escape in  both directions. By adjusting the ratio of
products escaping on the fuel  and oxidizer side, the investigation of
the effect of  bulk flow magnitude and direction  on a one-dimensional
unstrained diffusion flame is  made possible.  This configuration will
in particular allow  for the first time to create  a planar flame with
no net  flow across  the reaction  zone, i.e. a  flame where  all mass
transports are effected by diffusion only.

\section{Theoretical model} \addvspace{10pt}

We  start from  the  simplified theoretical  model  for the  idealized
chambered diffusion  flame of figure  \ref{fig:1DBurners}(a) (see e.g.
Cheatham   \&   Matalon   \cite{Cheatham2000}).    In   the   original
configuration, one  reactant (usually the  fuel) is supplied  from the
bottom  with a  uniform velocity  $\tilde  U$, while  the oxidizer  is
introduced  from the  top  and  reaches the  reaction  area solely  by
diffusion against the flow  of products.  The distribution of oxidizer
and products is `magically' kept uniform over the cross section by the
fast  top  stream  which  removes  combustion  products  and  supplies
reactant.  In the following, this model is adapted to the more general
symmetric  configuration shown  in figure  \ref{fig:1DBurners}(b) with
arbitrary bulk flow velocity and direction.

\subsection{Reaction-sheet approximation} \addvspace{10pt}

In   the  new   symmetric  burner   configuration,  shown   in  figure
\ref{fig:1DBurners}(b), the fuel  and oxidant are introduced uniformly
over  the burner  cross section  at $\tilde  x=0$ and  $\tilde  x= L$,
respectively.  In the  following, dimensional  variables  and unscaled
mass fractions $X$ are designated by \verb$~$. It is assumed that both
reactants follow  Fick's law  of mass diffusion  and burn in  a global
one-step irreversible reaction \ref{eq:Reaction}.

\begin{equation}
\nu_f \; \mbox{fuel} + \nu_x \; \mbox{oxidizer} \longrightarrow \;
\mbox{products} \; + Q
\label{eq:Reaction}
\end{equation}

Next, we make the drastic assumption of constant density $\tilde \rho$
and  transport properties  $D_i$, $\kappa$  and $C_p$,  independent of
temperature.   In  addition, it  is  reasonable  to  assume that  both
reactants  and  the  combustion  products  have  identical  $\bar  W$,
$\kappa$  and $C_p$  since both  reactants are  diluted with  an inert
($CO_2$) that constitutes the bulk of the mixture. As a consequence of
$\tilde  \rho  = \mbox{const.}$  the  bulk  flow  velocity $\tilde  U$
becomes constant  over the entire  burner length. This  simplifies the
steady-state  dimensional  equations of  conservation  of species  and
energy to:

\begin{equation}
  \tilde \rho  \tilde U \D{\tilde X_o}{\tilde x} - D_o \tilde \rho \DD{\tilde  X_o}{\tilde x}
   = -\nu_o W_o \tilde \omega
\label{eq:Species_o}
\end{equation}

\begin{equation}
  \tilde \rho \tilde U \D{\tilde X_f}{\tilde x} - D_f  \tilde \rho \DD{\tilde X_f}{\tilde x}
   = -\nu_f W_f \tilde \omega
\label{eq:Species_f}
\end{equation}

\begin{equation}
  \tilde \rho C_p  \tilde U \D{\tilde T}{\tilde x} - \lambda \DD{\tilde T}{\tilde x}  =  \tilde Q \tilde \omega
\label{eq:Energy}
\end{equation}

The  system is  made dimensionless  using  the chamber  length $L$  as
characteristic length and  $\kappa/L$ as characteristic velocity.  The
reference  temperature  is $\tilde  q/C_p$  where  $\tilde q=\tilde  Q
\tilde X_{f,L}/\nu_f  W_f$ represents the heat released  per unit mass
of fuel consumed. The mass fractions of fuel and oxidizer, finally are
scaled with  the conditions where  they are introduced: the  fuel mass
fraction by  $\tilde X_{f,0}$  at $\tilde x=0$  and the  oxidizer mass
fraction by $\phi \tilde X_{o,L}$ at $\tilde x=L$, where $\phi$ is the
fuel-to-oxidizer equivalence ratio.

\begin{equation}
  \phi  = \frac{\tilde  X_{f,L}/\nu_f  W_f}{\tilde  X_{o,0}/\nu_o
    W_o}
\label{eq:phi}
\end{equation}

Invoking  the  large  activation  energy approximation,  where  it  is
assumed  that no  reaction occurs  outside  of a  thin reaction  sheet
located  at $\tilde x  =\tilde x_f$,  the dimensionless  equations are
obtained as

\begin{equation}
U \D{X_o}{x} - Le_o^{-1} \DD{X_o}{x} = -\omega \delta(x-x_f)
 \label{eq:DL_Species_o}
\end{equation}

\begin{equation}
U \D{X_f}{x} - Le_f^{-1} \DD{X_f}{x} = -\omega \delta(x-x_f)
\label{eq:DL_Species_f}
\end{equation}

\begin{equation}
U \D{T}{x} - \DD{T}{x} = \omega \delta(x-x_f)
\label{eq:DL_Energy}
\end{equation}

where $\omega  \delta(x-x_f) = \nu_f  W_f L^2 \tilde  \omega(\tilde x)
[\kappa  \tilde  \rho   \tilde  X_{f,0}]^{-1}$.   The  non-dimensional
boundary conditions at the location  of fuel and oxidizer injection at
$x=0$ and $x=1$, respectively are

\begin{eqnarray}
x=0:& & \; X_f=1, X_o=0, T=T_0
\label{eq:Boundaries1}\\
x=1:& & \; X_o=\phi^{-1}, X_f=0, T=T_L \label{eq:Boundaries2}
\end{eqnarray}

To link  the solutions  across the flame  sheet at  $x=x_f$, equations
(\ref{eq:DL_Species_o})  -  (\ref{eq:DL_Energy})  are integrated  from
$x=x_f -0$ to $x=x_f +0$ to yield the jump conditions

\begin{eqnarray}
  \left[ \hspace{-1.5pt}\left[ T \right]\hspace{-1.5pt}\right]=\left[\hspace{-1.5pt}\left[ X_o
    \right]\hspace{-1.5pt}\right]  =  \left[\hspace{-1.5pt}\left[  X_f
    \right]\hspace{-1.5pt}\right]&=& 0\label{eq:Jump1}\\
  \left[\hspace{-3pt}\left[     \D{T}{x}     +     Le_o^{-1}\D{X_o}{x}
    \right]\hspace{-3pt}\right]&=& \nonumber \\ \left[\hspace{-3pt}\left[ \D{T}{x} +
      Le_f^{-1}\D{X_f}{x} \right]\hspace{-3pt}\right] &=& 0
\label{eq:Jump2}
\end{eqnarray}
where    the     operator    $\left[    \hspace{-1.5pt}\left[    \cdot
  \right]\hspace{-1.5pt}\right]$ represents the jump of the respective
quantity across  the reaction sheet (the difference  between its value
at  $x_f+0$ and  $x_f-0$). These  jump relations  (\ref{eq:Jump1}) and
(\ref{eq:Jump2})  represent respectively  continuity of  all variables
and  the fact  that the  reactants have  to reach  the flame  sheet in
stoichiometric  proportions  to  have  complete combustion,  i.e.   to
satisfy the boundary conditions $X_o(x=0)=0$ and $X_f(x=1)=0$.

\subsection{Model results} \addvspace{10pt}

The above  model corresponds  to the leading  order of  the activation
energy  expansion  of   Cheatham  \cite{Cheatham2000},  with  slightly
different  boundary  conditions  resulting  from  the  finite  chamber
length, and represents a  stable planar flame sheet. Solving equations
(\ref{eq:DL_Species_o})  -  (\ref{eq:DL_Energy})  in  each  sub-domain
$0\leq x \le x_f$ and $x_f \le x \leq 1$ and applying the boundary and
jump conditions  (\ref{eq:Boundaries1}) - (\ref{eq:Jump2})  yields the
flame  position $x_f$  given by  equation (\ref{eq:FlamePos})  and the
species   and   concentration   profiles  (\ref{eq:Sol_Species_o})   -
(\ref{eq:Sol_Energy}), where $T_f$  is the adiabatic flame temperature
defined by the jump relation.

\begin{equation}
\frac{e^{-ULe_fx_f}-1}{e^{ULe_o(1-x_f)}-1} = -\phi \label{eq:FlamePos}
\end{equation}

\begin{equation}
X_o = \begin{cases} 0 ~~ &\mbox{for} ~ 0\leq x \leq x_f\\
\frac{1}{\phi} \frac{e^{ULe_ox}-e^{ULe_ox_f}}{e^{ULe_o}-e^{ULe_ox_f}} ~~ &\mbox{for} ~ x_f \leq x \leq 1
\end{cases}
\label{eq:Sol_Species_o}
\end{equation}

\begin{equation}
X_f = \begin{cases} \frac{e^{ULe_fx}-e^{ULe_fx_f}}{1 - e^{ULe_fx_f}} ~~ &\mbox{for} ~ 0 \leq x \leq x_f\\
0 ~~ &\mbox{for} ~ x_f \leq x \leq 1
\end{cases}
\label{eq:Sol_Species_f}
\end{equation}

\begin{equation}
  T = \begin{cases} \frac{(T_f-T_0)e^{Ux} + T_0 e^{Ux_f} - T_f}{e^{Ux_f}
      - 1} ~~ &\mbox{for} ~ 0 \leq x \leq x_f\\
    \frac{(T_f - T_L)e^{Ux} + T_L e^{Ux_f} - T_f e^{U}}{e^{Ux_f}-e^{U}} ~~ &\mbox{for} ~ x_f \leq x \leq 1
\end{cases}
\label{eq:Sol_Energy}
\end{equation}

It  is interesting to  notice in  equation \ref{eq:FlamePos}  that the
flame  position in  the present  finite length  burner depends  on the
Lewis number of \emph{both}  species, in contrast to the semi-infinite
configuration of Cheatham \cite{Cheatham2000} where $x_f$ depends only
on the  Lewis number of the  species which has to  diffuse against the
bulk flow.

In the  limiting case  of vanishing bulk  flow $U=0$,  the expressions
(\ref{eq:FlamePos}) - (\ref{eq:Sol_Energy})  have to be evaluated with
H\^opital's rule to yield the limiting flame position
\begin{equation}
x_f(U=0) = \frac{\phi Le_o}{\phi Le_o+Le_f} \label{eq:FlamePos0}
\end{equation}
and linear concentration and temperature profiles not shown here.

\section{Experimental set-up} \addvspace{10pt}

The main  challenge when attempting to create  an unstrained chambered
flame is to supply the reactants and remove the products evenly across
the reaction area.  A novel way to address this  problem and produce a
truly  unstrained  flame has  recently  been successfully  implemented
\cite{LoJacono2005,  Robert2007}.  In  this first  implementation  the
reactant  which has  to diffuse  against the  bulk flow  is introduced
through an array of hypodermic  needles between which the products can
escape.

\subsection{Burner description} \addvspace{10pt}

\begin{figure}[ht!]
\centering
\includegraphics[width=67mm]{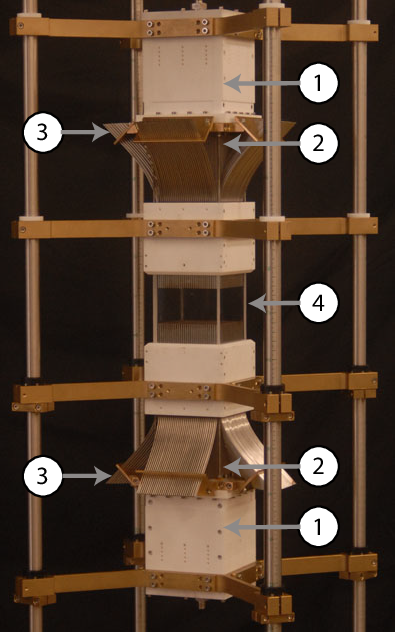}
\caption{Photograph of the partially assembled burner (without exhaust
  plenums and thermal insulation) with  all needle arrays in place. 1,
  Reactant  injection  plenums; 2,  Straight  injection needle  arrays
  emerging from  plenums; 3, Extraction needles  bent outwards between
  injection needles; 4, Quartz-walled reaction chamber.}
\label{fig:BurnerPhoto}
\end{figure}

In the present  paper we introduce the symmetric  `Mark II' version of
this  novel research  burner in  which both  reactants  are introduced
through needle arrays. A  photograph of the partially assembled burner
is shown  in figure \ref{fig:BurnerPhoto}.   The symmetric arrangement
of  the  product  extraction  manifolds exhausting  into  two  exhaust
plenums on both ends (not shown on the photograph) allows control over
both the magnitude and direction of the bulk flow through the flame by
adjusting the  pressure difference between the two  exhaust plenums. A
conceptual   sketch   of   the   burner   is   presented   in   figure
\ref{fig:1DBurners}(c).   The injection arrays  consist of  $31 \times
31=961$ stainless  steel hypodermic needles with an  outer diameter of
$1.2$ mm  and a wall  thickness of $0.1$  mm on a Cartesian  grid with
$2.5$  mm spacing.   Both arrays  are  introduced in  a quartz  walled
burning  chamber with  a  square cross-section  of  $77.5 \times  77.5
\mbox{ mm}^2$. The products are allowed to escape through a second set
of $32 \times  32 = 1024$ tubes of $1.2$ mm  I.D.  located between the
injection  needles.   These  extraction  tubes are  bent  outwards  to
deliver the exhaust gas into  annular exhaust plenums which are heated
to  prevent   condensation  in  the  exhaust   manifold  resulting  in
non-uniform product extraction. The  spacing between the two injection
arrays is adjustable up to $80$  mm, but for the present experiment it
was always kept at $40$ mm.

While the injection of reactant through a discrete array of hypodermic
needles is  inhomogeneous close to the  exit plane of  the needles, it
has  been  shown  experimentally  that this  inhomogeneity  disappears
beyond  an `injection  layer' of  thickness comparable  to  the needle
spacing.  In the `Mark I' version of the burner, the thickness of this
three-dimensional injection  layer has been measured  in the $1.5-2.5$
mm range \cite{Robert2007}, depending  on flow velocities.  Hence, the
desired  one-dimensional  reactant transport  is  realized beyond  the
injection layers.  For  all the results presented here,  the flame was
always   sufficiently  outside   of   the  injection   layers  to   be
one-dimensional  over nearly  the entire  burner cross  section.  Weak
inhomogeneities persist  near the edges  principally due to  heat loss
through the quartz windows of  the burning chamber which are necessary
to visualize  the flame structure. This inhomogeneity  over the burner
cross-section  is  clearly  visible  when approaching  extinction  and
results  in rapid cell  motion for  cellular flames  which complicates
measurements. In  the current version  of the burner, this  problem is
alleviated  by placing a  quartz tube  of $41$  mm I.D.,  spanning the
entire   chamber  length,   between  the   two  arrays   of  injection
needles. This creates two separate flames, a uniform one in the center
where the measurements are carried out and a surrounding annular flame
that experiences significant  gradients.  From the experiments carried
out  in the  Mark I  version of  the burner  one can  deduce  that the
residual  strain experienced  by  the central  flame  is below  $0.25$
s$^{-1}$ \cite{Robert2007}.

It is  however clear that  in our burner  configuration a part  of the
reactant injected against the bulk  flow turns around and is swept out
of  the  burner,  i.e.  does   not  diffuse  towards  the  flame.  The
determination  of the  effective species  transport towards  the flame
therefore requires mass spectrometry.

\subsection{Burner characterization} \addvspace{10pt}

Mixture composition  profiles taken  between the two  injection arrays
were  used to  define  the injection  layer  thickness, the  effective
boundary  conditions  and the  mixture  strength.   Sampling was  made
through a fused silica capillary (0.2 mm outer diameter) progressively
lowered through one of the  exhaust tubes. The capillary was protected
by  a 0.4  mm outer  diameter stainless  steel tube  to avoid  loss of
integrity as  the tip approaches the  flame. The sampling  rate of 0.1
ml/min was low  enough to induce only minimal  flow perturbations. The
samples  were  analyzed  with  a  quadrupole  mass  spectrometer  (MKS
Instruments, model  LM92 Cirrus).  The position of  the sampling point
was determined by photographs taken through the chamber windows.

\begin{figure}[ht!]
\centering
\psfrag{Xf}[bc]{\footnotesize $X_f$} %
\psfrag{x}[tc]{\large $x$} %
\psfrag{U}[bc]{$U$} %
\psfrag{Xo}[tc]{\footnotesize $X_o$} %
\psfrag{0}[cc]{\footnotesize $0$} %
\psfrag{0.2}[cc]{\footnotesize $0.2$} %
\psfrag{0.4}[cc]{\footnotesize $0.4$} %
\psfrag{0.6}[cc]{\footnotesize $0.6$} %
\psfrag{0.8}[cc]{\footnotesize $0.8$} %
\psfrag{1}[cc]{\footnotesize $1$} %
\includegraphics[width=67mm]{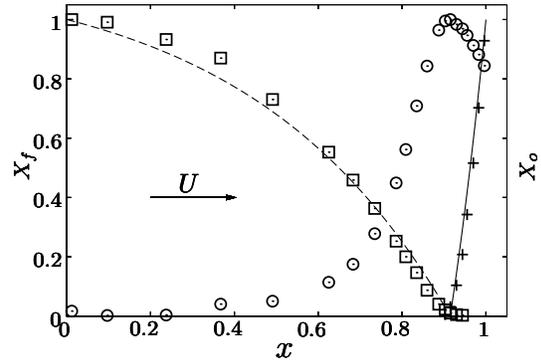}
\vspace{0.3cm}   \caption{Concentration  profiles  in   the  reaction
  chamber  with  fuel  advected  from  the left  (x=0)  for  $U=+7.05$
  ($\tilde U = 0.044$ m/s with $\tilde \rho$ and $\kappa$ evaluated at
  the adiabatic flame temperature  of $1550$ K), $\phi=1$, $Le_f=0.35$
  and $Le_o=0.86$.  $\Box$, experimental H$_2$ concentration $X_f$ ; - -
  -   -,  equation   \ref{eq:Sol_Species_f};  $+$,   experimental  O$_2$
  concentration $X_o$; -----, equation \ref{eq:Sol_Species_o}; $\circ$
  experimental H$_2$O concentration.}
\label{fig:ConcProfile}
\end{figure}

\begin{figure}[ht!]
\centering
\psfrag{Xf}[bc]{\footnotesize $X_f$} %
\psfrag{x}[tc]{\large $x$} %
\psfrag{U}[bc]{$U$} %
\psfrag{Xo}[tc]{\footnotesize $X_o$} %
\psfrag{0}[cc]{\footnotesize $0$} %
\psfrag{0.2}[cc]{\footnotesize $0.2$} %
\psfrag{0.4}[cc]{\footnotesize $0.4$} %
\psfrag{0.6}[cc]{\footnotesize $0.6$} %
\psfrag{0.8}[cc]{\footnotesize $0.8$} %
\psfrag{1}[cc]{\footnotesize $1$} %
\psfrag{0.5}[cc]{\footnotesize $0.5$} %
\psfrag{1.5}[cc]{\footnotesize $1.5$} %
\psfrag{2}[cc]{\footnotesize $2$} %
\includegraphics[width=67mm]{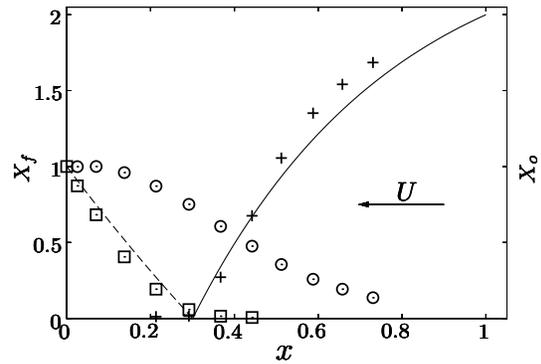}
\vspace{0.3cm}   \caption{Concentration  profiles  in   the  reaction
  chamber  with  fuel  counter-diffusing   from  the  left  (x=0)  for
  $U=-2.64$ ($\tilde U  = -0.036$ m/s with $\tilde  \rho$ and $\kappa$
  evaluated  at   the  adiabatic  flame  temperature   of  $1370$  K),
  $\phi=0.5$, $Le_f=0.25$ and $Le_o=0.80$.  Symbols are the same as in
  figure \ref{fig:ConcProfile}.}
\label{fig:ConcProfile_Rev}
\end{figure}

\begin{figure}[ht!]
\centering
\includegraphics[width=67mm]{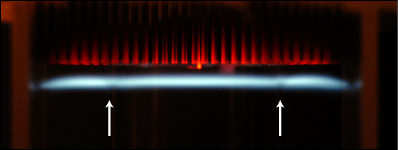}
\caption{Stable     flat     flame     corresponding     to     figure
  \ref{fig:ConcProfile}.   The fuel  is supplied  from the  bottom and
  advected  to the flame  by the  upward bulk  flow. The  white arrows
  indicate  the diameter  of the  inner  quartz tube  inserted in  the
  burner to separate  the central region, where the  flame is observed
  and probed, from the burner  edges.}
\label{fig:PhotoStableFlame}
\end{figure}

An elaborate calibration  procedure allows simultaneous measurement of
all four species  present within the burner (H2, H2O,  O2, CO2) over a
wide range  of mixture composition  using a single set  of calibration
data. The \emph{relative} error of these measurements is below 7\% for
the concentration range investigated  here.  The profiles presented in
figures \ref{fig:ConcProfile} and \ref{fig:ConcProfile_Rev} were taken
at the  center of the  burner cross section  in the fuel  advected and
fuel   counter-diffusing  configurations,   respectively.    They  are
compared  with the  results  expected from  the  simplified theory  of
Section 3.   The close agreement between  theoretical and experimental
concentration profiles demonstrates that  our experimental set-up is a
good realization  of a 1D  chambered diffusion flame. A  photograph of
the stable flat flame corresponding to figure \ref{fig:ConcProfile} is
shown  in figure  \ref{fig:PhotoStableFlame}.   It is  noted that  the
somewhat  higher luminosity of  the hypodermic  needles ends  near the
center as seen in figure \ref{fig:PhotoStableFlame} cannot be directly
associated with a nonuniform  temperature distribution.  The reason is
that  in the center,  the light  intensity is  a superposition  of the
``glows''  from 31  rows  of  needles, since  the  camera is  centered
relative to the burner and sees  right through the tube array.  On the
sides only the light emitted from  the first few rows reach the camera
because  of  the  wide  angle  objective used,  hence  leading  to  an
apparently brighter center.

\section{Experimental results} \addvspace{10pt}

\subsection{Flame position} \addvspace{10pt}

To measure the  flame location in the burner,  photographs where taken
from the side.  The images  where then analyzed and the flame position
determined as  the point  of maximum luminosity  in the center  of the
chamber.  This  optically determined flame location was  checked to be
in excellent  agreement with the  location of the  water concentration
and temperature maxima. For the sake of simplicity, the optical method
was used  for the  majority of the  results presented here.   For each
experimental condition, the  effective mixture composition immediately
next  to the  injection needle  array supplying  the counter-diffusing
species was measured by  mass spectrometry.  These local concentration
profiles, extending from the injection tube tips to the flame, allowed
the determination of  the effective boundary condition at  the edge of
the  downstream injection layer,  i.e.  the  amount of  reactant swept
into the  exhaust by  the bulk flow,  and hence the  effective mixture
strength in the  burner. Based on this concentration  profile, a point
beyond the injection layer was  chosen as the \emph{virtual origin} of
the   counter-diffusing  reactant.    The   mixture  composition   and
temperature at this point were then used as boundary condition and the
reduced chamber length as the  new reference length for the comparison
with the  simplified model of  section 3.  This procedure  allowed the
effective  mixture strength $\phi$  to be  kept constant,  despite the
varying  loss  of  counter-diffusing   species  into  the  exhaust  at
different bulk flow velocities.

\begin{figure}[ht!]
\centering
\psfrag{ylabel}[bc]{\large $x_f$} %
\psfrag{xlabel}[tc]{\footnotesize $U$} %
\psfrag{0}[cc]{\footnotesize $0$} %
\psfrag{0.2}[cc]{\footnotesize $0.2$} %
\psfrag{0.4}[cc]{\footnotesize $0.4$} %
\psfrag{0.6}[cc]{\footnotesize $0.6$} %
\psfrag{0.8}[cc]{\footnotesize $0.8$} %
\psfrag{1}[cc]{\footnotesize $1$} %
\psfrag{-20}[cc]{\footnotesize $-20$} %
\psfrag{-10}[cc]{\footnotesize $-10$} %
\psfrag{10}[cc]{\footnotesize $10$} %
\psfrag{20}[cc]{\footnotesize $20$} %
\psfrag{30}[cc]{\footnotesize $30$} %
\psfrag{40}[cc]{\footnotesize $40$} %
\includegraphics[width=67mm]{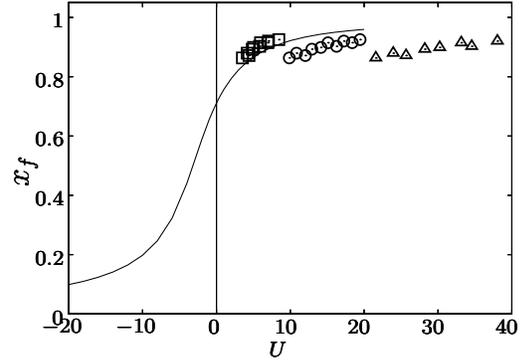}
\vspace{0.3cm}
\caption{Effect of bulk  flow velocity on flame position  for the case
  of    counter-diffusing    oxidant.     The   $\kappa$    used    to
  non-dimensionalize  the  bulk velocity  is  evaluated  at the  flame
  ($\Box$),   at   the   advecting   inlet  ($\circ$)   and   at   the
  counter-diffusing inlet ($\triangle$). ------, theoretical $x_f$ for
  $\phi=1$, $Le_f=0.35$ and $Le_o=0.85$.}
\label{fig:BulkFlowPhi1}
\end{figure}

\begin{figure}[ht!]
\centering
\psfrag{ylabel}[bc]{\large $x_f$} %
\psfrag{xlabel}[tc]{\footnotesize $U$} %
\psfrag{0}[cc]{\footnotesize $0$} %
\psfrag{0.2}[cc]{\footnotesize $0.2$} %
\psfrag{0.4}[cc]{\footnotesize $0.4$} %
\psfrag{0.6}[cc]{\footnotesize $0.6$} %
\psfrag{0.8}[cc]{\footnotesize $0.8$} %
\psfrag{1}[cc]{\footnotesize $1$} %
\psfrag{-20}[cc]{\footnotesize $-20$} %
\psfrag{-15}[cc]{\footnotesize $-15$} %
\psfrag{-10}[cc]{\footnotesize $-10$} %
\psfrag{-5}[cc]{\footnotesize $-5$} %
\psfrag{5}[cc]{\footnotesize $5$} %
\psfrag{10}[cc]{\footnotesize $10$} %
\includegraphics[width=67mm]{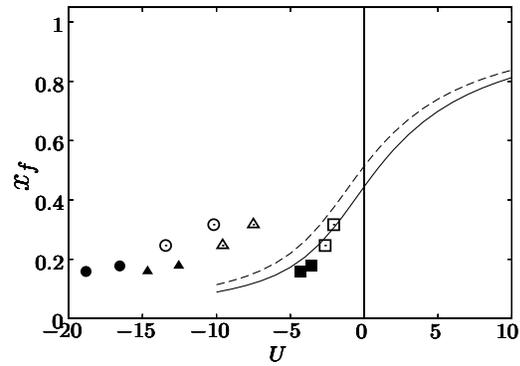}
\vspace{0.3cm}
\caption{Effect  of   bulk  flow  velocity  on  flame
  position for  the case  of counter-diffusing fuel  with $\phi=0.25$,
  $Le_f=0.25$ and $Le_o=0.80$ (solid symbols and theoretical line) and
  $\phi=0.33$,  $Le_f=0.25$ and $Le_o=0.80$  (open symbols  and dashed
  theoretical line). The $\kappa$  used to non-dimensionalize the bulk
  velocity  is evaluated  at the  flame ($\Box,\blacksquare$),  at the
  advecting inlet ($\circ,\bullet$) and at the counter-diffusing inlet
  ($\triangle,\blacktriangle$).}
\label{fig:BulkFlowPhi3325}
\end{figure}

A  comparison between  the measured  flame position  and  the position
calculated  from equation  \ref{eq:FlamePos} is  presented  in figures
\ref{fig:BulkFlowPhi1} and \ref{fig:BulkFlowPhi3325} and clearly shows
the influence of  bulk flow magnitude and direction.   To evaluate the
theoretical  flame position,  the  experimental Lewis  numbers of  the
reactant mixtures  were taken  at the advecting  and counter-diffusing
inlets, respectively.  The dimensional experimental  bulk velocity was
determined from the mass flow and the mixture density evaluated at the
upstream combustion chamber inlet temperature. The thermal diffusivity
$\kappa$, to non-dimensionalize the  bulk velocity, on the other hand,
was  evaluated   at  three  different  temperatures:   at  both  inlet
temperatures and  at the  adiabatic flame temperature  determined with
the Cantera software package \cite{Goodwin2003, GoodwinXXXX}.

These results  show good agreement  between the simplified  theory for
the flame  position (equation \ref{eq:FlamePos})  and the experimental
results. It  clearly appears that  the adiabatic flame  temperature is
the  most  appropriate  temperature   to  evaluate  $\kappa$  for  the
non-dimensionalization of  the bulk  velocity.  The figures  also show
that the flame position is more accurately predicted by the simplified
theory when the oxygen rather than the hydrogen is counter-diffusing.

Several factors can be invoked  to explain the differences between the
two data  sets.  The assumption of constant  transport properties made
in  the  theoretical  development  is  not  equally  drastic  in  both
configurations.  Equation  (\ref{eq:FlamePos}) reveals that  the flame
position   is  mainly   determined  by   the  Lewis   number   of  the
counter-diffusing species.  In  the present experiments, $Le_o$ varied
only slightly from  the injection point to the  flame sheet (from 0.861
to 0.865)  when the oxidant was  counter diffusing. When  the fuel was
counter-diffusing, the $Le_f$  variation was more significant, between
0.25 at the injection and 0.22 at the flame, on average.  In addition,
the  thermal  expansion,  also  neglected in  the  simplified  theory,
decreases  the  transport  of  the counter-diffusing  species  to  the
reaction  zone. This effect  increases the  effective $\phi$  when the
oxidant  is   counter-diffusing  and  decreases  it   in  the  reverse
configuration (note that a decreasing $\phi$ lowers the curve $x_f(U)$
in figures \ref{fig:BulkFlowPhi1} and \ref{fig:BulkFlowPhi3325}).

\subsection{Flame stability} \addvspace{10pt}

\begin{figure}[ht!]
\centering
\includegraphics[width=67mm]{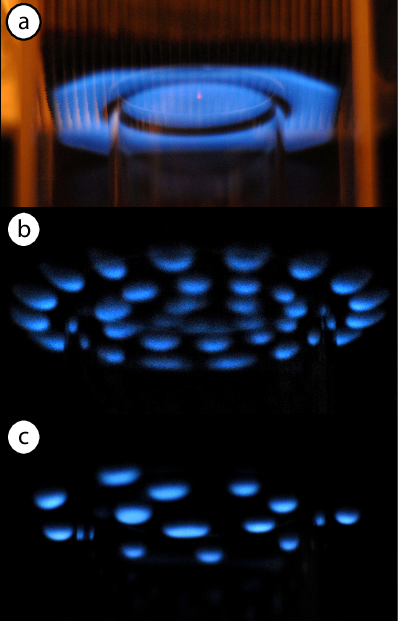}
\caption{Low angle  view ($-30^\circ$) of  the burner used  to capture
  the flame  structure. a)  Reference view of  stable flame,  with the
  central quartz tube visible; (b), cellular flame with advecting fuel
  ($\tilde U = 0.036$ m/s); (c), cellular flame with counter-diffusing
  fuel ($\tilde U = -0.026$ m/s).}
\label{fig:UnstableFlames}
\end{figure}

The stable  flame sheet  is rendered unstable  by slowly  lowering the
hydrogen concentration  in the fuel stream. When  approaching the lean
extinction  limit, the stable  flame sheet  fragments into  cells. The
asymptotic    theory   of    diffusion    flames   \cite{Cheatham2000,
  Metzener2006}  predicts that  a  cellular flame  pattern forms  near
extinction  if the  reactants  Lewis numbers  are sufficiently  small.
Examples    of     cellular    flames    are     shown    in    figure
\ref{fig:UnstableFlames}  for  the   two  cases  of  counter-diffusing
oxidant and fuel.   As mentioned previously, a quartz  tube of $41$ mm
I.D., seen in figure (\ref{fig:UnstableFlames}a), was inserted between
the two injection arrays to separate the central region with extremely
low strain  from the edges  of the burner where  temperature gradients
resulting  from   heat  loss  through   the  burner  walls   are  more
significant.

The introduction  of the  quartz tube in  the chamber allows  for easy
measurement of  the cell size.  In the advected fuel  configuration of
figure (\ref{fig:UnstableFlames}b), there are $11$ cells aligned along
the  inner tube perimeter,  yielding a  cell size  of about  $9.1$ mm.
When   the   oxidant  is   the   advected   species,   as  in   figure
(\ref{fig:UnstableFlames}c),  only $6$  cells are  visible  yielding a
wave length of about $14$ mm.

The scaling of the cell  size has been addressed in stability analyzes
of cellular diffusion  flames \cite{Cheatham2000, Kim1996}. To compare
the  two  cellular flames  of  figure (\ref{fig:UnstableFlames}),  the
problem  of where to  evaluate the  transport properties  arises. Here
they are  assumed equal in the  two configurations since  at the flame
the  mixture consists  only of  inert and  products,  with vanishingly
small concentrations  of reactants.  In  both of the  references cited
above, the  cell size  is found  to scale with  $1/\tilde U$  when all
mixture properties are held  constant. Considering the (drastic) model
assumptions,  this scaling  appears  to be  confirmed  by the  present
preliminary experiments which produce a  cell size ratio of $1.55$ for
a ratio  of bulk  velocity magnitude of  $1.4$. The evaluation  of the
scaling of cell size with  reactant properties, i.e. the evaluation of
the  pre-factor  of   $1/\tilde  U$,  on  the  other   hand,  and  the
investigation of a wider range  of bulk velocities including the limit
$\tilde U \to 0$ will be the subject of future experiments.

\section{Conclusions} \addvspace{10pt}

The  burner  presented  in  this  paper has  evolved  from  the  first
experimental   realization   of  a   1D   chambered  diffusion   flame
\cite{LoJacono2005,  Robert2007}.  The novel  symmetric design  of the
reactant supply to the burning chamber allows the investigation of the
effect of  bulk flow magnitude and  direction on a  diffusion flame in
the absence of strain. This design is in particular aimed at the first
investigation of a purely diffusive  unstrained flame in the zero bulk
flow  limit.  The  practical difficulty  with this  design due  to the
unknown proportion of  fuel and/or oxidant lost into  the exhausts has
been overcome by the systematic use of mass spectroscopy in the burner
chamber. The  key to  the successful use  of the mass  spectrometer in
this situation  with large  concentration variations is  an innovative
calibration   procedure  \cite{Robert2007}  which   yields  continuous
concentration  profiles of  the 4  species present  in the  burner and
permits  the determination of  the effective  mixture strength  of the
flame.

The  good agreement  between the  present experiments  and  the simple
theory for  the flame position and the  species concentration profiles
demonstrates that our new symmetric  burner is a good approximation of
a  chambered   one-dimensional  diffusion  flame,   with  one  species
transported  against  the bulk  flow  by  diffusion only.  Preliminary
observations of cellular flame patterns for both the advected fuel and
counter-diffusing fuel configurations are  compatible with a cell size
scaling proportional to $1/\tilde U$ when all the transport properties
are held constant.

The authors wish  to express their gratitude for  the support received
from the Swiss National Science Foundation under Grant 20020-108074.


 \footnotesize
 \baselineskip 9pt

\end{document}